\documentclass[fleqn,twoside]{article}
	
\usepackage{insa}
\usepackage{graphicx}
\title{Search and study of Quark Gluon Plasma at the CERN-LHC}

\author{Tapan Nayak\address{Variable Energy Cyclotron Centre, Kolkata 700064} and 
Bikash Sinha\address{Variable Energy Cyclotron Centre, Kolkata 700064 and Saha Institute of 
Nuclear Physics, Kolkata 700064} }

\begin{document}
\thispagestyle{empty}
\begin{abstract}

The major aim of nucleus-nucleus collisions at the LHC is to study the 
physics of strongly interacting matter and the quark gluon plasma (QGP), 
formed in extreme conditions of temperature and energy density. We give a 
brief overview of the experimental program and discuss the signatures and 
observables for a detailed study of QGP matter.

\end{abstract}

\maketitle

\section{Introduction}

The Large Hadron Collider (LHC) at CERN is designed to deliver colliding 
proton-proton~(\mbox{p-p}) beams 
at center-of-mass energies of 14~TeV and lead-lead~(\mbox{Pb-Pb})
beams at 5.5~A TeV. Collisions at these
unprecedented energies offer outstanding opportunities for new physics in compliance
with the standard model and beyond. The \mbox{p-p} collisions offer the tantalizing
possibility of discovering the Higgs Boson, the missing link in the standard model. The
collisions of two lead nuclei, on the other hand, 
will create a speck of very high temperature and high 
energy density matter, called the quark gluon plasma (QGP), where the properties of the 
system are 
governed by the quarks and gluons. 
According to conventional wisdom, the Universe is believed to have been in this state
only a few microseconds after the Big Bang. So the LHC is going to have a ``peep''
into the very early stages of the creation of the Universe, and, of course, into the history of
its evolution through space and time. A review of the recent status of QGP may be found in 
reference~\cite{QM09}.

The LHC is going to answer several questions of fundamental interest. The ones which
concern us in this report are the two most novel features of quantum chromodynamics (QCD), 
{\it viz.}, the asymptotic freedom and quark confinement. Discovered in 1973, the asymptotic 
freedom tells that within the nucleons, quarks move mostly as free non-interacting particles.
This has earned Gross, Wilczek and Politzer the Nobel Prize in physics in 2004. It is also known
that color-charged particles, such as quarks, are confined within 
hadrons. Statistical QCD calculations which take into account these properties predict that
strongly interacting systems at high temperature and/or energy density are composed of weakly 
interacting quarks and gluons. Such a phase consisting of (almost) free quarks and gluons
is termed as the quark gluon plasma.
In 1974, Prof. T.D. Lee realized that it would be interesting to explore
the phenomenon of QGP formation by distributing high energy and high density over a
relatively large volume. Heavy-ion collisions at relativistic energies offer such
a possibility of creating high energy and high density matter in the laboratory.
Soon after, dedicated experimental programs have started to search and study 
the QGP matter in great detail by colliding heavy-ions at dedicated accelerator facilities.
The heavy-ion program at CERN-LHC is the latest in this endeavor.

Heavy-ion physics is an integral part of the baseline program of the LHC.
The major aim of the experimental program is to identify and access 
most of the QGP signatures for a detailed study of the QGP properties.
The focus is to study how collective phenomena and microscopic properties, involving
many degrees of freedom, emerge from microscopic laws of elementary particle physics. 
In addition, with the LHC heavy-ions, one can access a novel range of Bjorken-$x$ values where 
strong nuclear shadowing is expected.  
The initial density of the low-x gluons, accessible at LHC energies, is expected to be close to 
saturation.
With these studies, the LHC will turn out to be a discovery machine
for various types of new physics and will explore QCD phenomena in great detail.

\section{QCD Phase Diagram}

\begin{figure}[hbt]
\begin{center}
\includegraphics[width=8.1cm, height=5.5cm]{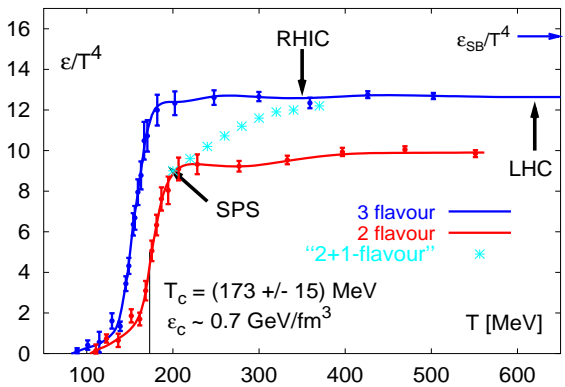}
\caption{\label{fig:lattice_figure}Lattice QCD calculations for energy density as a 
function of temperature.}
\end{center}
\end{figure}

While embarking on the detailed experimental program, it is essential
to know the conditions required to create the QGP phase. 
Lattice QCD calculations have been performed by formulating QCD on the lattice and 
performing numerical Monte Carlo simulations \cite{phil,sourendu}. Necessary
conditions for the QGP phase transition can be obtained from these calculations.
Results of such a calculation \cite{karsch} is shown in 
Figure~\ref{fig:lattice_figure} for energy density ($\epsilon$) as a function of temperature ($T$). 
The energy density is seen to exhibit the typical behaviour
of a system with phase transition, where an abrupt change is observed within 
a very narrow temperature range
around the critical temperature, $T_C$. These calculations give
a critical temperature, $T_C \sim 173\pm 15$~MeV corresponding to the critical
energy density of $\epsilon_C \sim 0.7$~GeV/fm$^3$.

\begin{figure}[hbt]
\begin{center}
\includegraphics[width=8.1cm, height=6.1cm]{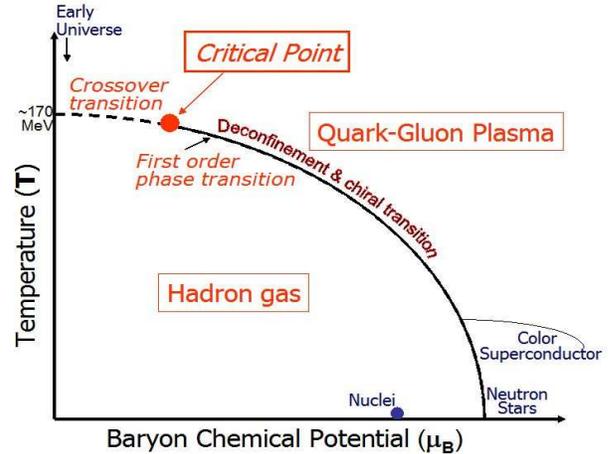}
\caption{\label{fig:qgp_phase} A schematic of the QCD phase diagram of nuclear matter.}
\end{center}
\end{figure}

A schematic of the QCD phase diagram is shown in Figure~\ref{fig:qgp_phase}, as a function of the
temperature (T) and the baryon chemical potential $\mu_{\rm B}$.
This diagram signifies the separation of the
QGP to hadronic phase of matter. Lattice QCD models predict a smooth crossover
at high T and small $\mu_{\rm B}$ while there are expectations for a first order
transition at smaller T and larger $\mu_{\rm B}$. 
The existence of the critical point
has also been predicted where a sharp transition between the QGP phase and the hadronic
phase first appears. The exact location of the critical point is not known yet,
but various calculations suggest that it might be within the reach of heavy-ion experiments.

\section{Experimental program for QGP search and the LHC}

Dedicated programs to create and study the QGP phase have started in early eighties
with collisions of heavy-ions at relativistic energies. A tremendous amount of effort has been
put for the development of four generations of experiments. Table~1 gives the list
of some of the heavy-ion facilities. The quest for the search and study of QGP started first 
with the Au beam at 1~A.GeV at the Bevalac in Berkeley. The early success of the experiments in terms
of bringing out the collective nature of the produced matter prompted the scientists
at Brookhaven National Laboratory (BNL) 
and CERN to make concrete programs for the future
accelerator developments for heavy ions. The next milestone came with the acceleration
of Au beam at 11.7~A.GeV at the BNL-AGS and Pb beam at 158~A.GeV at the CERN-SPS.
First hints of the formation of a new state of matter has been obtained from
the SPS data \cite{sps,bedanga_van_hove}. The Relativistic Heavy Ion Collider (RHIC) 
started becoming operational
in the year 2000 with \mbox{Au-Au} collisions at $\sqrt{s_{\rm NN}}=130$~GeV and
soon after to top  \mbox{Au-Au} energies of $\sqrt{s_{\rm NN}}=200$~GeV. The
experimental program at RHIC included four experiments, two large and two small with
the involvement of more than 1200 physicists. At present, the RHIC experiments bring 
out highest quality data from \mbox{p-p}, \mbox{d-Au}, \mbox{Cu-Cu} and \mbox{Au-Au} at various
energies. Strong evidence for the production of extreme hot and dense matter has
been seen. The matter formed at RHIC has been termed as 
sQGP (strongly coupled QGP) \cite{RHIC}. The
RHIC results, in combination with the ones from AGS and SPS, have enhanced our
understandings of the QCD matter at different temperatures and densities.

\begin{table}[hbt]
{\offinterlineskip \tabskip=14pt\halign{ \hfil #\hfil & #\hfil & #\hfil \tabskip=0pt \cr
\hline
\vspace{.1cm}
Laboratory   &   Energy (A.GeV)       &   $\Delta$y \cr \hline
\vspace{.1cm}
LBL-Bevelac  & 2.0 Fixed target       &   1.81 \cr
\vspace{.1cm}
Dubna        & 4.1 Fixed target       &   2.36 \cr
\vspace{.1cm}
BNL-AGS      & 11.7 (Au) Fixed target &   3.4  \cr
\vspace{.1cm}
CERN-SPS     & 158 (Pb) Fixed target  &   6.0  \cr
\vspace{.1cm}
BNL-RHIC     & 200 (Au+Au) Collider   & 11.7   \cr
\vspace{.1cm}
CERN-LHC     & 5500 (Pb+Pb) Collider  & 18.0   \cr
\hline
}}
\vspace{.2cm}
\caption{ \it Accelerator facilities for heavy-ions. Only top energies for given facilities are listed. The last column gives the accessible rapidity range.}
\end{table}

\begin{figure}[hbt]
\begin{center}
\includegraphics[scale=0.385]{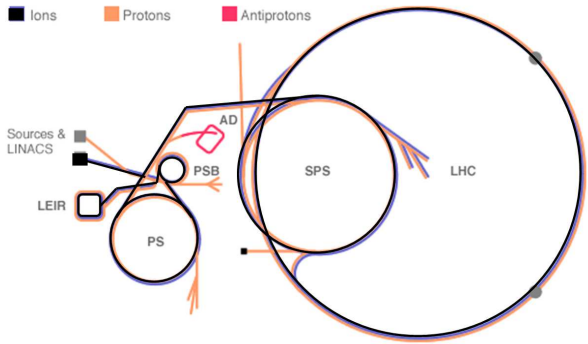}
\caption{\label{fig:LHC_ring}The CERN accelerator complex. The ions pass through the LIER, then to the
PS, the SPS and finally to the LHC.}
\end{center}
\end{figure}

In October 1990, in a workshop held at Aachen, Germany, the then CERN Director-General
Carlo Rubbia, while discussing the case for the proposed Large Hadron Collider, emphasized
the idea of providing \mbox{p-p} collisions as well as heavy-ions \cite{CERN_c}.
In the CERN accelerator complex (see Figure~\ref{fig:LHC_ring}),
the ions are passed into the Low Energy Ion Ring (LEIR), then passed to the PS, the SPS and
finally to the LHC. 

The ALICE experiment at the LHC is designed specifically for heavy-ion physics~\cite{fabjan}.
The other two major experiments, ATLAS and CMS, also have heavy-ion programs~\cite{ent,grau}.
Both of these experiments will focus more on jets, photons and muon channels.
The typical expected yearly running times of LHC are 
of the order of $10^7$seconds for \mbox{p-p} collisions and $10^6$seconds for 
heavier systems~\cite{antinori}. Table~2 gives the center-of-mass energy and 
expected luminosity at LHC for some typical collision systems~\cite{antinori,godbole}. 
Due to the limited rate capability of the ALICE detector, the luminosity for \mbox{p-p} 
collisions will be limited to a maximum of 
$10^{31}$cm$^{-2}$s$^{-1}$ for the ALICE experiment.


\begin{table}[hbt]
{\offinterlineskip \tabskip=14pt\halign{ \hfil #\hfil & #\hfil & #\hfil & #\hfil \tabskip=0pt \cr
\hline
\vspace{.1cm}
System       & $\sqrt{s_{\rm NN}}$(TeV) & L$_0$(cm$^{-2}$s$^{-1}$) & $\sigma_{\rm geom}$(b) \cr \hline
\vspace{.1cm}
pp                    &  14.0    &   10$^{34}$             &   ~0.1         \cr
\vspace{.1cm}
PbPb                  &   5.5    &   10$^{27}$             &   7.7          \cr
\vspace{.1cm}
pPb                   &   8.8    &   10$^{29}$             &   1.9          \cr
\vspace{.1cm}
ArAr                  &   6.3    &   10$^{29}$             &   2.7          \cr
\hline
}}
\vspace{.2cm}
\caption{\it Luminosities expected for different collision systems with different center-of-mass 
energies \cite{antinori}.}
\end{table}

A comparison of some of the basic parameters measured at the SPS and RHIC with
expected values for the LHC is given in 
Table~3. The table lists the psedorapidity density of charged
particles (d$N_{ch}$/d$\eta$), formation time of QGP 
($\tau_0$), the energy density ($\epsilon$) for $\tau_0=1$, the initial temperature
in terms of the critical temperature ($T_{\rm C}$),
QGP life time ($\tau_{QGP}$) and the freezeout time 
($\tau_{f}$) for the top center-of-mass energies ($\sqrt{s_{\rm NN}}$)
of these accelerators. At the LHC, 
with the increase of beam energy to about 28 times more compared 
to that of
RHIC, the number of charged particles from the produced fireball 
increase by more than three times, the formation time of the QGP
decreases significantly, the matter becomes much hotter, denser and long lived.
All these conditions are conducive to opening up new physics domain at the LHC.



\begin{table}[hbt]
{\offinterlineskip \tabskip=14pt\halign{ \hfil #\hfil & #\hfil & #\hfil & #\hfil \tabskip=0pt \cr
\hline
\vspace{.1cm}
Condition                           &   SPS     &   RHIC            &    LHC          \cr \hline
\vspace{.1cm}
$\sqrt{s_{\rm NN}}$ (GeV)           &   17.3    &   200             &   5500          \cr
\vspace{.1cm} 
d$N_{ch}$/d$\eta$                   &    450    &   600             &   $1200 - 4000$ \cr
\vspace{.1cm}
$\tau_0$ (fm/$c$)                   & $\sim$1   &   $\sim$0.2       &  $\sim$0.1      \cr
\vspace{.1cm}
$\epsilon$(GeV/fm$^3$)              & 2.5       &   4-5             & 10-40           \cr
\vspace{.1cm}
$T/T_{\rm C}$                       & 1.1       &   1.9             &   $3 - 4$       \cr
\vspace{.1cm}
$\tau_{QGP}$ (fm/c)                 & $\sim$1   &   $2-4$           &  $>$4           \cr
\vspace{.1cm}
$\tau_{f}$ (fm/c)                   & $\sim$10  &   $\sim$20        &  $\sim$30       \cr
\hline
}}
\vspace{.2cm}
\caption{\it Conditions created in central heavy-ion collisions at the top energies of SPS, RHIC and projections for LHC.}
\end{table}

\section{The ALICE Experiment at LHC}

\begin{figure}[hbt]
\begin{center}
\includegraphics[scale=0.4]{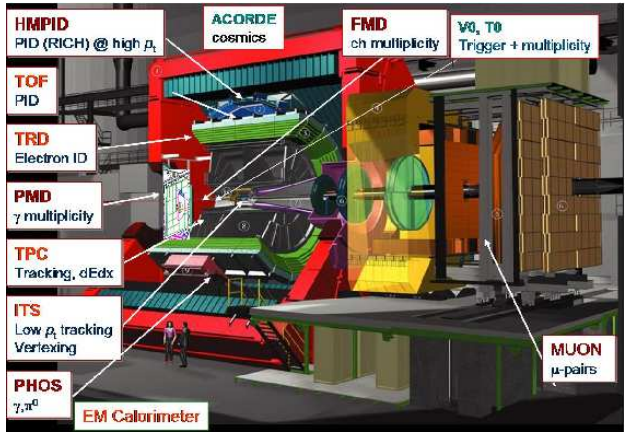}
\caption{\label{fig:ALICE}The experimental setup for the ALICE experiment 
\cite{alice-ppr-1,alice-ppr-2} 
at the LHC. }
\end{center}
\end{figure}

\begin{figure}[hbt]
\begin{center}
\includegraphics[scale=0.34]{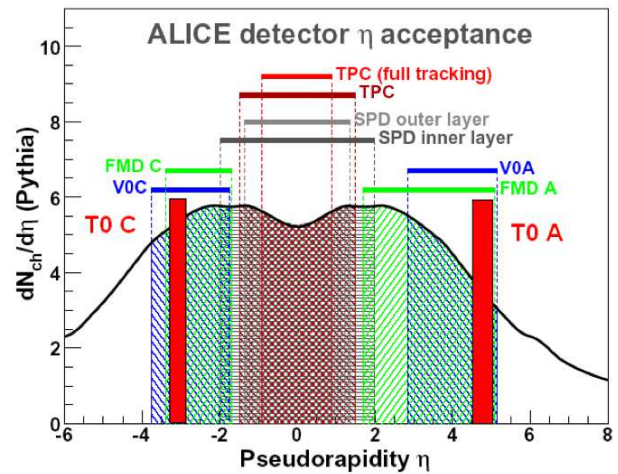}
\caption{\label{fig:coverage} Coverages of various ALICE detectors for
charged particle measurements. Superimposed is the Pythia event generator
prediction for \mbox{p-p} collisions at 14~TeV. }
\end{center}
\end{figure}

The experimental setup of the ALICE experiment \cite{alice-ppr-1,alice-ppr-2}
is shown in Figure~\ref{fig:ALICE}.
The detector setup can be broadly described by three groups of detectors:
the central barrel, the forward detectors and the
muon spectrometer \cite{fabjan}. Coverages of various detectors for charged particle 
measurements is shown in Figure~\ref{fig:coverage}. 


\subsection{The central barrel}
\medskip
The central barrel consists of the Inner Tracking System (ITS),
Time Projection Chamber (TPC), Transition Radiation Detector (TRD),
the Time of Flight (TOF) detector and the newly proposed electromagnetic calorimeter.
The design goal is to have low material budget and low magnetic field (B$\le$0.5T)
in order to be sensitive to low-$p_T$~ particles.

The ITS consists of six layers of silicon detectors.
The two innermost layers consist of silicon pixel
detectors (which covers  up to the $|\eta|<1.98$),
followed by two layers of silicon drift detectors and the last two
outer layers are of silicon strip detectors.
The ITS is designed to handle high particle density
expected in \mbox{Pb-Pb} collisions.
The main uses of ITS will be to (i) determine the primary vertex of the collision,
(ii) localize the secondary vertices for decays of hyperons and D and
B mesons, (iii) track and identify particles with momentum below 200~MeV/$c$,
(iv) improve the tracking in the central barrel and (v) provide minimum bias
and high multiplicity triggers.

The TPC is the main tracking detector of ALICE.
It is cylindrical
in shape. The active volume has an inner radius of about 85~cm, an
outer radius of about 250~cm, and an overall length along the beam
direction of 500~cm. With this the coverage of TPC becomes $|\eta|<0.9$.
The TPC is made of large cylindrical field cage,
filled
with $90~{\rm m}^3$ of Ne/CO$_{2}$/N$_2$ (90/10/5), in which
the primary electrons are transported over a distance of up to 2.5~m on
either side of the central electrode to the end plates. Multi-wire
proportional chambers with cathode pad readout are mounted into 18
trapezoidal sectors at each end plate. The TPC is the primary device
for obtaining charged-particle momentum measurements and particle
identification. It also provides an independent measure of primary vertex.
The combination of ITS and a large Time Projection Chamber (TPC) provides powerful tracking
with excellent momentum resolutions (about 2\% to 5\%) 
from 100~MeV/{\it c} to 100~GeV/{\it c}.

The central barrel is equipped with a 
Transition Radiation Detector (TRD) for electron identification
above 1~GeV/$c$, where the pion rejection
capability from TPC is no longer sufficient. The TRD, in combination with other
central barrel detectors, will provide sufficient electron identification to measure
the production of light and heavy vector meson resonances and the dilepton
continuum produced in \mbox{p-p} and \mbox{Pb-Pb} collisions.
It consists of 540 individual readout detector modules arranged
in 18 super modules. Each module consists of a radiator of 4.8~cm thicknes,
a multi-wire proportional readout chamber along with front-end elctronics.
The gas mixture in the readout chamber is Xe/CO$_2$ (85\%/15\%). 

The particle identification of hadrons in the intermediate momentum
range (to about 4~GeV/$c$ depending on the particle species)
is improved significantly with the inclusion of a 
Time-Of-Flight (TOF) system. The TOF system consists of 
Multi-gap Resistive-Plate Chambers (MRPC). The key aspect of these chambers is that the
electric field is high and uniform over the whole sensitive gaseous volume of the detector.
Inoziation produced by traversing charged particle starts a gas avalanche process
which eventually generates the observed signals on the pick-up electrodes.
The intrinsic time resolution of 40ps has been achieved with an efficiency of close to 100\%.

The central arm includes a  High-Momentum Particle Identification Detector
(HMPID) for the identification of hadrons at $p_{\rm t}$~$>$1~GeV/$c$.
The detector is based on proximity-focusing Ring Imaging Cherenkov
(RICH) counters and consists of seven modules of about 1.5$\times$1.5~m$^2$ each, mounted in
an independent support cradle. The coverage of HMPID is limited ($-0.6<\eta<0.6$  with 
$1.2^\circ < \phi < 58.8^\circ$.

The measurements of  low $p_T$ direct photons and high-$p_T$ $\pi^0$  
are performed by a single arm high resolution photon (elctromagnetic) spectrometer (PHOS),
consisting of lead-tungstate ({\rm PbWO}$_4$) crystals. The major requirements of the PHOS
include the ability to identify photons, discriminate direct photons from decay photons and
perform momentum measurements over a wide dynamic range with
high energy and spatial measurements. PHOS covers approximately a quarter of a unit
of pseudorapidity, $-0.12\le \eta \le 0.12$, and 100$^\circ$ in azimuthal angle. The
total area is $\sim$8m$^2$.

A new electromagnetic calorimeter (EMCAL) is being planned which will improve 
ALICE capability for measurement of high energy jets.

\subsection{The Forward detectors}
\medskip
The ALICE experiment is equipped with a set of forward detectors, such as the
a Forward Multiplicity Detector (FMD), Photon Multiplicity Detector (PMD),
Zero Degree Calorimeters (ZDC),  and detectors for trigger and timing (V0, T0). 

The FMD, consisting of several rings of silicon detectors,
covers a very large range in pseudo-rapidity ($-3.4<\eta<-1.7$ and $1.7<\eta<5.0$).
The main function of FMD is to provide charged-particle multiplicity information.

The V0 detector consists of two arrays (each array with 32 segments) 
of scintillator counters, called V0A and V0C,
which are installed on both sides of the ALICE interaction point. 
With a time resolution of about 600ps the detector provides minimum-bias triggers 
for in pp and A--A collisions. V0 detector helps in monitoring of the beam luminosity
and reducing the beam-gas contributions.

The T0 detector consists of two arrays of Cherenkov counters (12 counters
per array) and has an excellent time resolution (about 50ps). The main purpose of
T0 detector is to provide start time (T0) for the TOF detector. T0 detector also measures
the vertex posision and performs several trigger related functions.

The Photon Multiplicity Detector (PMD) in ALICE has been conceived, designed and fabricated
by an Indian collaboration comprising of Variable Energy Cyclotron Centre, Kolkata,
Institute of Physics, Bhubaneswar, Indian Institute of Technology, Mumbai,
Rajasthan University, Panjab University, and Jammu Univerisity.
The PMD is designed to measure the  multiplicity and 
spatial distribution of photons in the forward rapidity region of   $2.1\le \eta \le 3.9$. 
The photon measurements along with those of charged particles from the Forward Multiplicity Detector 
(FMD) provide vital information in terms of the limiting fragmentation, the equation of state of matter 
from the determination of elliptic flow, information about phase transition and the formation of 
disoriented chiral condensates (DCC). 

\begin{figure}[hbt]
\begin{center}
\includegraphics[scale=0.43]{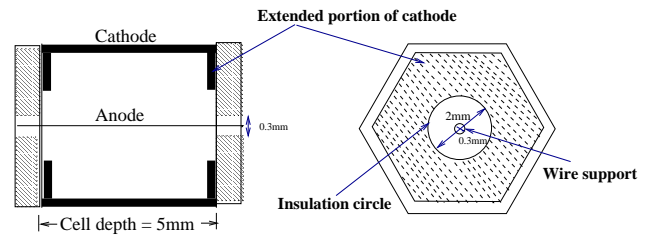}
\caption{\label{fig:cell_cs}Schematic diagram of the cross section of a unit cell of the PMD in the 
ALICE experiment.} 
\end{center}
\end{figure}

\begin{figure}[hbt]
\begin{center}
\includegraphics[scale=0.4]{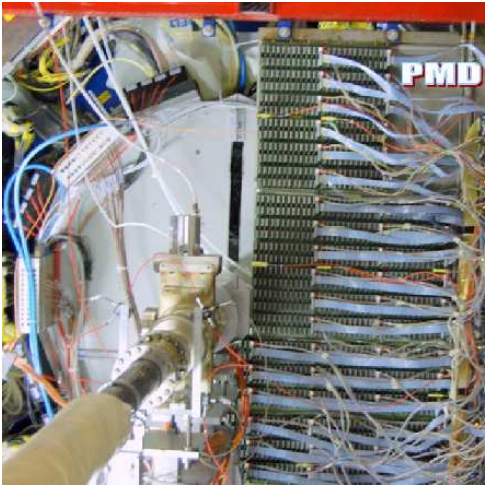}
\caption{\label{fig:PMD} Photograph of the partially installed PMD in the ALICE cavern.}
\end{center}
\end{figure}

Because of the large particle density in the forward region,
calorimetric techniques for photon measurements are not feasible.
The PMD uses the preshower detector, where a three radiation length thick lead converter is sandwiched 
between two planes of highly granular gas proportional counters. The information from one of the gas 
detector planes, placed in front of the converter is used to veto the charged particles, whereas the 
preshower data from the other detector plane is used for photon identification. 
The granularity and the converter thickness of the PMD are optimized for
high particle density so that the overlap of photon showers is minimal.
The PMD consists of 
221184 honeycomb shaped proportional counters, each of 0.25cm$^2$ area.
Each counter has a honeycomb shaped cathode extended towards a
20~$\mu$m thick gold-plated tungsten wire kept at a ground potential
at the centre of each cell.
The schematic diagram of the unit cell is shown in Figure~\ref{fig:cell_cs}.
The optimal operating voltage for the detector
is $-1400$~Volts which forms part of the plateau region of the proportional zone. 
The efficiency is about
96\% for charged pions at this voltage.

\begin{figure}[hbt]
\begin{center}
\includegraphics[scale=1.5]{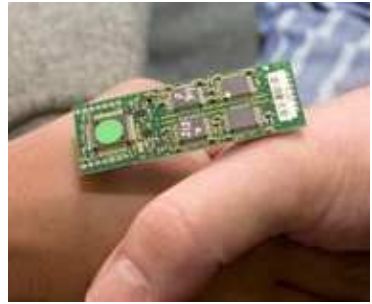}
\caption{\label{fig:MANAS} A closeup view of the electronic readout board of ALICE dimuon 
spectrometer showing four MANAS chips. }
\end{center}
\end{figure}

The PMD is assembled in two equal halves. Each half has independent cooling, gas supply
and electronics accessories. A partial installation with one half of the PMD in the
ALICE cavern is shown in Figure~\ref{fig:PMD}.
The front-end electronics consists 
of MANAS (Multiplexed ANAlog Signal Processor) chips (Figure~\ref{fig:MANAS}) 
for anode signal processing and the Cluster Read 
Out Concentrator Unit System (CROCUS) for data acquisition.

\begin{figure}[hbt]
\begin{center}
\includegraphics[scale=0.7]{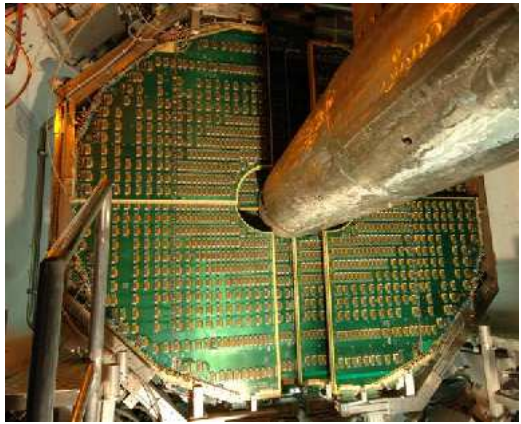}
\caption{\label{fig:DIMUON} Photograph of the second station of dimuon spectrometer. }
\end{center}
\end{figure}

\subsection{The Dimuon Spectrometer}
\medskip

The ALICE forward muon spectrometer will study the complete spectrum of heavy quarkonia 
($J/\psi$, $\psi'$, $\Upsilon$, $\Upsilon'$, $\Upsilon''$) via their decay in the $\mu^+\mu^-$ channel.
The spectrometer acceptance covers the pseudorapidity interval $2\le \eta \le 4$ and 
the resonances can be detected down to zero transverse momentum. The invariant mass resolution 
is of the order of 70~MeV in the $J/\psi$ region. The spectrometer consists of a front absorber 
($\sim 10 \lambda_{int}$) which absorbs hadrons and photons from the interaction vertex, a large
dipole magnet, a highly granular tracking system and a trigger system placed behind a passive muon
filter wall.

The dipole magnet is positioned at about 7m from the interaction vertex. The magnetic field
(Bnom=0.7T, 3Tm field integral) is what defined by the requirements of mass resolution.

The muon tracking system is based on low thickness cathode pad chambers. 
The chambers are arranged in five
stations (each of 2 chambers), 
two of the stations placed before, one inside and two after the dipole. To keep the occupancy at
the 5\% level, a high segmentaion of the readout pads is needed, leading to a total of about one
million channels. The chambers have a position resolution of about 70 microns in the bending
direction and 800 microns in the non-bending direction. 

The frontend pulse processing of all the channels
is done by MANAS chip which include preamplifier, a shaper and a multiplexer.  
A closeup view of the electronic readout board showing four MANAS chips 
is shown in Figure~\ref{fig:MANAS}. Saha Institute of Nuclear Physics, Kolkata
is instrumental in the design and fabrication of the MANAS chip.

The station two muon tracking spectrometer has been designed and built in India by the
the Saha Institute and 
Aligarh Muslim University. A photograph of the second muon station is shown in 
Figure~\ref{fig:DIMUON}. The detector has been successfully installed in the ALICE cavern.

The trigger system is designed to select heavy quark resonance decays. The selection is made
on the transverse momentum on individual muons. Four planes of resistive plate chambers (RPCs)
arranged in two stations and positioned behind a passive muon filter provide the transverse
momentum of each muon.

\section{Signals and observables}

The QGP phase manifests itself in several different forms. The signals are not quite unique and one
needs to make a comprehensive study of all available probes in order to make any firm conclusion.
The signals and observables can be categorized in terms of global observables, event-by-event 
fluctuations, DCC search, electromagnetic probes, heavy quarks and quarkonia, and to the
physics at high $p_T$ and jets. Finally possibilities at low Bjorken-$x$ values will also tell us about
the possibility of color glass condensate picture \cite{larry}. A good volume on the
predictions for \mbox{Pb-Pb} collisions at  $\sqrt{s_{\rm NN}}=5.5$~TeV may be found in
\cite{LHC_predictions}. Below we review some of the accessible QGP signatures for the LHC experiments.

\medskip
\noindent
{\large{\bf Global observables }}
\medskip

The comprehensive study of global observables provides valuable information
for thermal and chemical analysis of the freeze-out conditions. Some of these observables 
include the 
rapidity distributions of charged and identified particles, momentum spectra, particle ratios, 
flow and the size of the fireball.

\begin{figure}
\begin{center}
\includegraphics[scale=0.61]{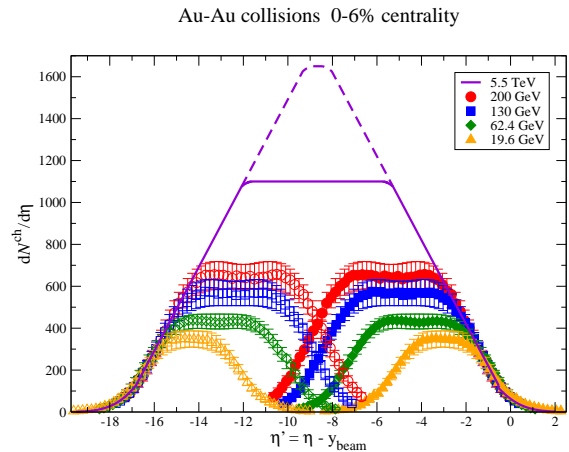}
\end{center}
\caption{\label{fig:lhc_dndeta} Charged particle multiplicity density scaled by their
beam rapidities and extrapolation to LHC energies \cite{urs}. }
\end{figure}

Particle multiplicity
measurements will constitute one of the first measurements in ALICE. This will be an
eagerly awaited result during the first days of the LHC startup.
There are large uncertainties in theoretical predictions \cite{LHC_predictions} 
for rapidity density at central
rapidity (the rapidity, $y$, is defined in terms of energy and longitudinal momentum; 
alternatively one uses pseudorapidity, $\eta$, which is related to the angle of emission ($\theta$)
of the particle). 
Figure~\ref{fig:lhc_dndeta} shows the masured pseudorapidity distributions (shifted by the beam
rapidity, $y_{\rm beam}$) of charged particles for several beam energies at RHIC.
An extrapolation \cite{urs} to \mbox{Pb-Pb} collisions at LHC energies
may provide a good estimation. The extrapolated
results, shown in Figure~\ref{fig:lhc_dndeta}, gives a value of $\sim$1100 at mid rapidity.
Other theoretical estimations come up with numbers between 1200-2500.
The ALICE detector is optimized for the charged-particle density of 4000
and its performance is checked up to a value of 8000. Thus ALICE will be able to handle
the large multiplicity data from LHC. 

\begin{figure}
\begin{center}
\vspace*{-0.5cm}
\includegraphics[width=8.1cm, height=6.1cm]{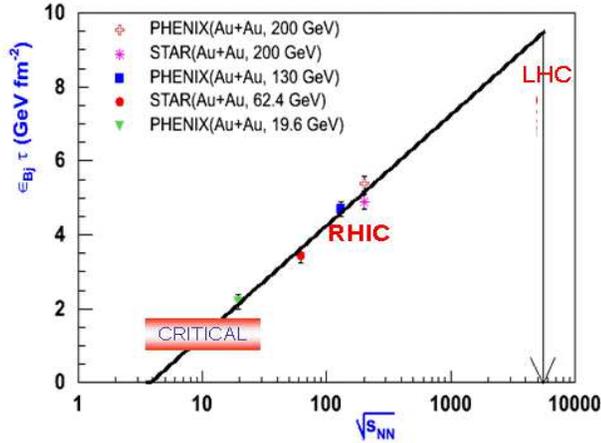}
\end{center}
\caption{\label{fig:Bjorken} Energy density as a function of beam energy. The figure
indicates the possible location of the critical point. A simple
extrapolation is also made to the LHC energy \cite{raghu}, but the actual value is
expected to be much higher.}
\end{figure}

Momentum spectra of charged particles and identified
hadrons will constitute the next sets of measurements from ALICE. Fitting these spectra with
specific models allow one to extract mean transverse momenta, temperature, radial flow
and other observables \cite{LHC_predictions}.

The data on the charged particle multiplicity and mean transverse momenta
allow one to get an estimation of the energy density. Figure~\ref{fig:Bjorken} gives \cite{raghu}
the energy densities as a function of beam energy. The energy density where the critical point
may occur is marked in the figure. The extrapolated value to the LHC shows the energy density to be
about two times larger compared to the values at RHIC.

\begin{figure}
\begin{center}
\includegraphics[scale=0.55]{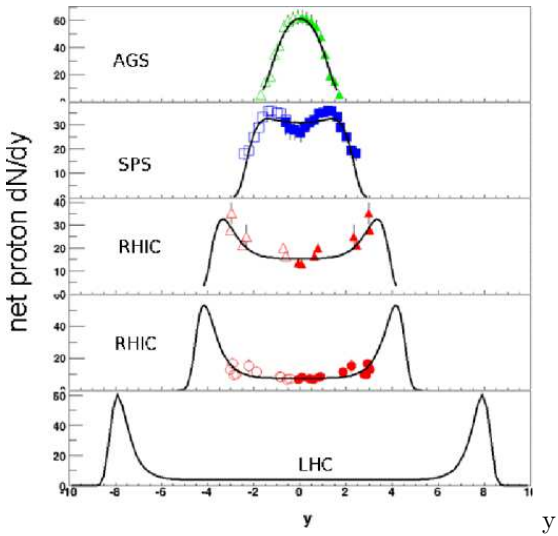}
y\end{center}
\caption{\label{fig:net_proton} Net proton rapidity distributions at AGS, SPS and RHIC. Plots for two different energies at RHIC are shown. 
Extrapolation to the LHC energy shows a complete transparency for a 
large rapidity range \cite{esumi}.}
\end{figure}

\begin{figure}[hbt]
\begin{center}
\includegraphics[width=8.1cm, height=4.7cm]{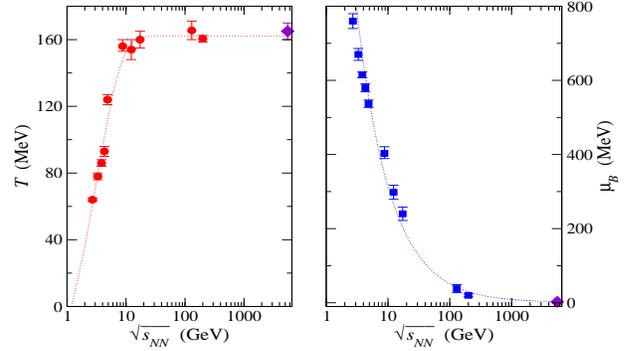}
\end{center}
\caption{\label{fig:temp_mu}Estimation of freeze-out temperature and chemical
potential from thermal model fits as a function of centre-of-mass energy of the
collision \cite{urs}.}
\end{figure}

The measured net proton rapidity density distributions for AGS, SPS, RHIC energies 
\cite{esumi}
are shown in Figure \ref{fig:net_proton} with extrapolations to LHC energies. 
At AGS energies the number of produced antiproton is very small and the net-proton
distribution is similar to the proton distribution. At SPS and higher energies 
the net proton rapidity distribution shows double hump with a dip around $y=0$. 
This shows that beyond the SPS energy the reaction is beginning to be transparent
in the sense that fewer of the original baryons are found at midrapidity after
the collisions. A complete transparency can be expected at LHC energies for a 
large rapidity range.

\begin{figure}[hbt]
\begin{center}
\includegraphics[width=7.4cm, height=5.4cm]{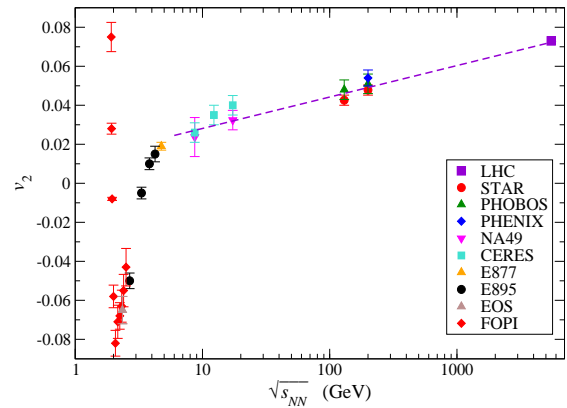}
\end{center}
\caption{\label{fig:lhc_v2} 
Elliptic flow, $v_2$ as a function of centre-of-mass energy of the
collision \cite{urs} for existing data and extrapolated to LHC.
}
\end{figure}

Estimation of freeze-out temperature and chemical potential
is essential for mapping out the QGP phase space.
Thermal model fits have been used \cite{urs} to estimate these values from the
measured spectra and particle ratios.
This is shown in Figure~\ref{fig:temp_mu}. The
chemical potential at top RHIC energies is between 20-40~MeV and at LHC energies it
is expected to quite low, less than 10~MeV. 

An important measure of the collective dynamics of heavy-ion collisions is the elliptic
flow ($v_2$). Figure~\ref{fig:lhc_v2} shows excitation function of $v_2$ for mid-central
collisions. Because of the large values of $v_2$ at RHIC energies, in agreement with
the value for an ideal fluid, the formation of a perfect liquid is ascertained 
at RHIC energies. A simple extrapolation of the $v_2$ value has been made
for mid-central collisions at the LHC.

\begin{figure}[hbt]
\begin{center}
\includegraphics[width=8.1cm, height=10.0cm]{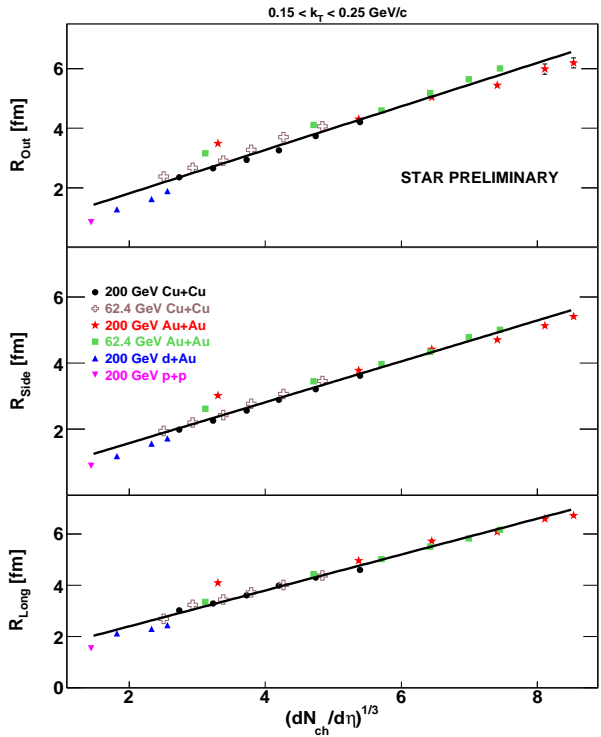}
\end{center}
\caption{\label{fig:hbt} Pion source radii dependence on charged particle
multiplicity. The lines are plotted to guide the eye and represent
linear fits to the data \cite{das}. Extrapolating to the LHC energies
would mean a very large source size at the freezeout.}
\end{figure}

The dynamical evolution of the collision fireball and its
space-time structure has been traditionally studied using two-particle (HBT)
correlations. The multiplicity and transverse momentum dependence of three-dimensional 
pion interferometric radii in Au-Au and Cu-Cu collisions at different RHIC
energies \cite{das} have been shown in Figure~{\ref{fig:hbt}.
The freezeout volume estimates with charged pions measured from such studies,
show linear dependence as a function of charge particle multiplicity
indicating consistent behaviors with a universal mean-free-path at freeze-out.
It will be interesting to find out what happens at the LHC as the fireball is supposed to 
be much larger and longer lived. At LHC energies, because of the large number
of produced particles, it may be possible to extract HBT radii on an event by event. 

\bigskip
\noindent
{\large{\bf Event-by-event Physics}}
\medskip

It is expected that the hot and dense system created in heavy-ion collisions at
LHC energies will show very characteristic behaviour of QGP phase transition, which
may vary dramatically from one event to the other. As the number of particles produced
at these energies is quite large, it is possible to study various observables in every
single event and study their event-by-event fluctuations. the correlation and fluctuation
measures provide possible ways to study variations in physical quantities from event to event.

Fluctuations of thermodynamic quantities such as temperature and entropy have been proposed
to give evidence for the existence of QGP phase transition and also provide direct insight
into the properties of the system created in heavy-ion collisions \cite{alice-ppr-2,nayak}.
Active study of event-by-event fluctuations in heavy-ion collisions was initiated by
experiments at the SPS and now very much applied to the data at RHIC energies. 
These studies include fluctuations in particle multiplicity,
particle ratios, net charge, mean transverse momentum and formation of DCC domains.
In order to be more sensitive to the origin of fluctuations, differential
measures have been adopted where the analysis is performed at different scales (varying bins
in $\eta$ and $\phi$). 

For \mbox{p-p} collisions, the soft and semihard parts of the multi-particle
production are successfully described in terms of colour strings stretched between
the projectile and target. For nuclear collisions the number of strings grows with the
growing energy and the number of participants in the collision. One has to take
into account the interaction between strings in the form of their fusion and 
percolation \cite{pajares}. Long range correlations were proposed as the main tool to
study these phenomenon. The method of
long-range correlation coefficients for different colliding systems and centralities will help to
understand the critical fluctuation relevant to the string fusion and percolation phenomena.

The capabilities of the ALICE detector can
be explored \cite{alice-ppr-2} in terms of measurements of temperature
and $p_T$ fluctuations, multiplicity and strangeness fluctuations, fluctuations of conserved
quantities (such as net-charge and net-baryon), balance functions, fluctuation in azimuthal
anisotropy, fluctuation in space-time parameters and long range correlations.
The effect of minijets and jets on the event-by-event studies will have to be clearly
understood in order to make any conclusion. This is important in order to make any inference
about the nature of event-by-event fluctuation as well as to understand the effect
of jets passing through the high density medium created in heavy-ion collisions.

\bigskip
\noindent
{\large{\bf Disoriented Chiral Condensates}}
\medskip

The QCD phase transition is predicted to be accompanied by chiral symmetry restoration
at high temperature and densities. It leads to the formation of a chiral condensate in an extended
domain such that the direction of the condensate is misaligned from that of the true vacuum.
This disoriented chiral condensate (DCC) \cite{dcc} results in the production of low momentum pions in
a single isospin direction, leading to large fluctuations in the ratio ($f$) 
of neutral to charged pions. Normally a distribution of $f$ would follow a binomial form with
a mean of 1/3, whereas within a DCC domain it takes the form, $1/2\sqrt{f}$. This 
feature of the $f$ distribution can be used to characterize a DCC domain.

The formation of DCC has been hypothesized in the context of explaining observed abnormal
events from cosmic ray experiments which had either excess of charged-particles (Centauro 
events) or excess of neutrals (anti-Centauro events) \cite{cosmic}. 
DCC search has been carried out by the miniMax experiment at Fermilab where no evidence was
found \cite{minimax}. A thorough search in \mbox{Pb-Pb} collisions at 
$\sqrt{s_{\rm NN}}$=17.3~GeV
was performed by the WA98 collaboration at CERN-SPS based on correlation study of photon and
charged-particle multiplicities \cite{wa98-dcc-1,wa98-dcc-2} using data from the preshower
photon multiplicity detector (PMD)
and silicon pad multiplicity detector, respectively. A detailed analysis of centrality and
acceptance dependence of multiplicity fluctuations had shown absence of any significance 
non-statistical fluctuations. Using the results from experimental data, mixed events and models
which incorporate DCC domain, an upper limit on DCC production has been set. This is shown in
Figure~\ref{fig:dcc}.

\begin{figure}[hbt]
\begin{center}
\includegraphics[width=7.9cm, height=10.4cm]{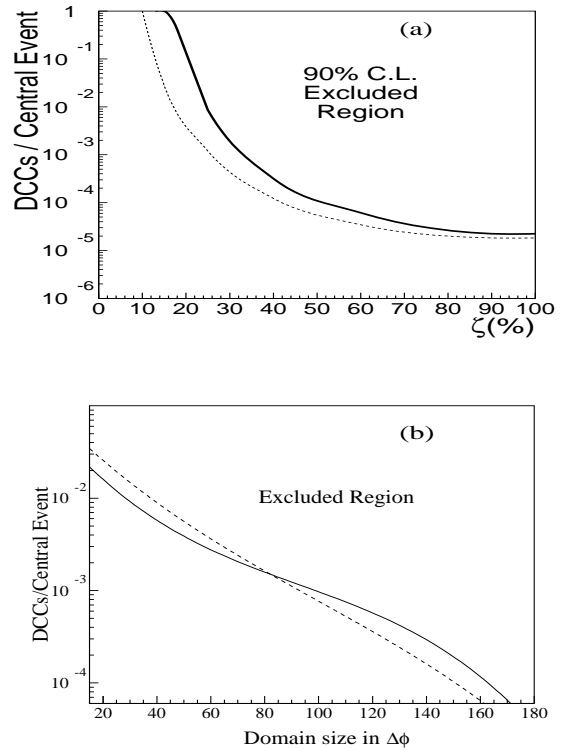}
\end{center}
\caption{\label{fig:dcc} 
Upper limit (90\% confidence level) 
of DCC production at SPS energies based on fluctuations of charged particles and photons.
Panel (a) shows the limit as a function of the fraction of DCC pions \cite{wa98-dcc-1}, 
whereas panel (b)
gives it as a function of DCC domain size in azimuthal angle \cite{wa98-dcc-2}.
The solid and dashed lines in lower panel correspond two centrality bins: top $5\%$ and 
$5-10\%$ of the minimum bias cross section.}
\end{figure}

The search for DCC at LHC energies will be possible by the ALICE experiment where the
production of charged particles and photons will be studied in both the central and
forward rapidity regions. In the central rapidity, a combination of the PHOS detector and
TPC along with TOF will be used. In the forward rapidity region, PMD and FMD combination
in a common coverage of $2.1 \le \eta \le 3.9$ will be used.
The detection of DCC will mainly be done through the
event-by-event fluctuation study \cite{dcc_probe}.
Theoretical calculations also suggest that DCC formation
might give rise to other signatures such as enhanced correlation of kaons and
enhanced production of baryons, particularly, $\Omega$ and $\bar\Omega$.
In addition, DCC formation may be seen through the HBT correlations of identified pions and
non-equilibrium photons in the sample of direct photons. Various analysis methods,
adopted for DCC search in ALICE, may be found in \cite{alice-ppr-2}.

\bigskip
\noindent
{\large{\bf Electromagnetic probes}}
\medskip

Interaction of the charged particles produced in the nucleus-nucleus collisions
emit real and virtual photons (lepton pairs). Owing to the nature of the interaction 
they undergo minimal scatterings and are by far the best markers of the entire 
space-time evolution of the collision \cite{dinesh,mclerran,gale,alam1,alam2}.

The extraction of direct photons from experimental data is complicated by the huge
amount of background from hadronic decays.
The WA80 collaboration \cite{wa80} provided the first interesting result with 
a $p_T$ dependent
upper limit on the direct photon production in \mbox{S-Au} collisions at lab energy of
200~A.GeV which was supported by theoretical calculations \cite{srivastava1}.  
The high quality single photon data obtained from \mbox{Pb-Pb} collisions at CERN-SPS by the 
WA98 Collaboration \cite{wa98-photon} have been the focus of considerable 
interest \cite{alam3}. 
The direct photon spectra at low $p_T$ from the PHENIX experiment \cite{photon_phenix}
is seen to be consistent with rates calculated with thermal photon emission taken into account.
The photon spectrum in ALICE will be measured with the PHOS spectrometer. 
The direct photons will be identified as an excess of photons when compared with the
decay photons. The systematic error is expected to be about 8\% \cite{alice-ppr-2}.

A better insight into the nature of the evolving system can be provided by
the HBT correlations of direct photons \cite{bass_photon_hbt,alam5}.
These photons are emitted during all stages of the collision and serve as a deep probe of the
hot and dense matter.
Although photon HBT has been suggested quite some time back, 
because of the difficulty in measurement (mainly because of low production
cross section and large background) it took quite some years to finally have first
results by the WA98 collaboration \cite{wa98_photon_hbt} and 
the final RHIC results \cite{das_photon_hbt} are expected soon.
In ALICE direct photon HBT will be possible by the use of the PHOS detector.

The dilepton production has been studied since the start of the QGP program.
This is because the process of dilepton production is sensitive at low dilepton masses
to possible chiral restoration in dense matter, at intermediate masses to thermal dilepton
production from the QGP and at higher masses it provides the best access to charmonium production 
and its predicted in-medium modification. At low masses $M<1$ GeV, thermal dilepton production
is mediated by broad vector meson, $\rho(770)$ in the hadronic phase. $\rho(770)$ 
is very short lived (lifetime of only 1.3fm/$c$) and has a strong coupling to
$\pi\pi$ channel.  The `in-medium' modifications of $\rho(770)$
mass and width close to the QCD phase boundary have long been considered as the prime signature for
chiral symmetry restoration. Thermal dileptons have been studied at SPS~\cite{damj} and
RHIC~\cite{averbeck} energies.
The di-electron spectra, measured by the PHENIX
experiment shows that the data are in good agreement with the
cocktail of hadron decay sources over the full mass range. 
Di-electrons will be measured in ALICE with the central tracking system for masses above
0.2~GeV/c$^2$. Details of measurement and studies in ALICE may be found in \cite{alice-ppr-2}. 

\begin{figure}
\begin{center}
\includegraphics[width=8.1cm, height=6.1cm]{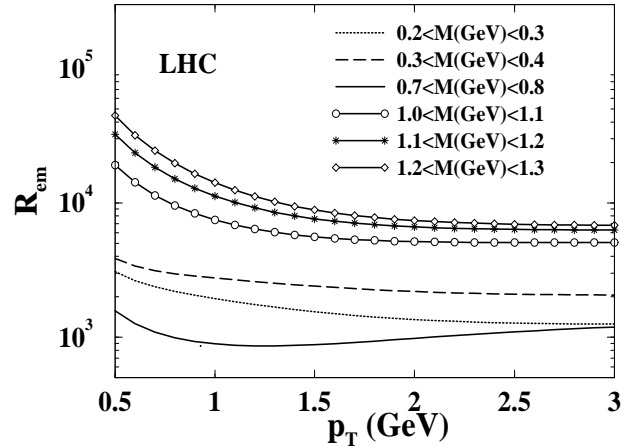}
\end{center}
\caption{\label{fig:ratio} The thermal photon to dilepton ratio at LHC energies 
as a function of transverse
momentum, $p_T$, for various invariant mass windows \cite{alam4}.}
\end{figure}

It has been proposed since long back to look at the ratio of 
transverse momentum distribution of thermal photons to dileptons \cite{sinha}.
Recently it has been shown \cite{alam4} that this ratio reaches a plateau beyond a certain
value of transverse momentum as shown in Figure \ref{fig:ratio}. 
The ratio is sensitive to initial conditions and thus can be
used to estimate the initial temperature of the system by properly selecting
the transverse momentum and invariance mass windows.

\bigskip
\noindent
{\large{\bf Heavy flavours and Quarkonia}}
\medskip

The study of quarkonia production represents one of the most powerful methods
to probe the medium of the fireball. Heavy quarkonia are sensitive to the collision
dynamics at both short and long timescales and are expected to be sensitive to
plasma formation.
Suppressions of $J/\Psi$ and $\Psi'$ (bound states of charmonia) have long been
considered to be the most direct signatures of QGP formation \cite{matsui}. 
Indeed such suppressions have been observed in \mbox{Pb-Pb} collisions at the SPS energies
\cite{NA50}. 
Recent high accuracy measurement by the NA60 collaboration \cite{NA60}
for \mbox{In-In} collisions
also shows such a suppression. There have been contrasting calculations which explain
the suppressions even without invoking a scenario of QGP formation.
Higher statistics data from RHIC experiments \cite{jpsiphenix,jpsistar} 
are awaited for better distinctions between 
model predictions. 

Several models predictions exist~\cite{LHC_predictions}
at LHC energies for production of heavy quarks and quarkonia
in heavy-ion collisions. 
At the LHC, the much higher energy offers the possibility of measuring
quarkonia production with significant statistics. 
ALICE will measure heavy quarkonia in both di-electron and dimuon channels. 
One of the main methods of physics analysis will be made by normalizing quarkonia yields in 
heavy-ion collisions to reference processes, such as the production in pp collisions.
The suppression of quarkonia, if any, will be clearly seen in these analyses.

\bigskip
\bigskip
\noindent
{\large{\bf High $p_T$ and jets }}
\medskip

Jets are defined in QCD as cascades of consecutive emissions of partons in a narrow cone
initiated by partons from an initial hard scattering. Because of QCD confinement, particles 
carrying a color charge, such as quarks, cannot exist in free form. Therefore they fragment 
into hadrons before they can be directly detected, becoming jets. The measurement of jets
help to determine the properties of the original quark and give indications
of the properties of the hot and dense medium. High $p_T$ partons produced in the initial stage
of a nucleus-nucleus collision are expected to undergo multiple interactions inside the 
collision region prior to hadronization. Thus the energy of the partons is reduced through
collisional energy loss and medium induced gluon radiation. This so called jet-quenching
mechanism has been postulated as a tool to probe the properties of the QGP phase of matter.
The measurement of $\gamma$-jet events is expected to provide an unique 
probe of parton energy loss \cite{srivastava,renk}. 

At RHIC, the study of jets have been performed by introducing a nuclear modification
factor and also by the method of back-to-back correlations.
The nuclear modification factor, $R_{\rm AA}$ has been defined as as the jet-yield
within a given $p_T$ bin for nucleus-nucleus collisions, normalized to corresponding yield for
\mbox{p-p} collisions. The quantity, $R_{\rm AA}$, is defined in such a way that it becomes
unity if the nucleus-nucleus collision were a mere superposition of number of nucleon-nucleon
collisions. At RHIC energies, strong suppressions are observed in central \mbox{Au-Au}
collisions corresponding to \mbox{p-p} \cite{phenixjet,starjet}.
The back-to-back correlation strength of high $p_{\rm T}$ hadrons is seen to be
sensitive to the in-medium path length of the parton. The study by the STAR 
experiment over a broad range in transverse momenta \cite{star_dijet} 
for \mbox{d-Au} and  \mbox{Au-Au} collisions show that
whereas there is no significant change in the near side
peak with the increase of $p_{\rm T}$ of the associated particles, 
the away-side correlation strength decreases 
from \mbox{d-Au} to central \mbox{Au-Au} collisions. 
The strongest modifications of the correlated yields are seen at lower associated $p_{\rm T}$. 

At the LHC, the copious production of jets will enable us to understand the medium 
properties in a very detailed manner. Details of energy loss mechanism and predictions for
LHC energies may be found in \cite{LHC_predictions}. Within the ALICE acceptance it is 
possible to map out the energy dependence of jet fragmentation over a wide kinematic
range, up to a transverse energy of about 200~GeV \cite{alice-ppr-2}. Major problem
in jet reconstruction in nucleus-nucleus collisions comes from the large background
of the underlying event. Addition of the new electromagnetic calorimeter in ALICE
is expected to improve the jet-energy resolution and thereby jet reconstruction capability.

\bigskip
\noindent
{\large{\bf QCD at small $x$ and forward physics }}
\medskip

At large collision energy and relatively low momentum transfer ($Q$), one expects a new
regime of QCD known as saturation \cite{larry,gelis}. This is described in a picture of
colour glass condensate (CGC) where a saturation scale emerges naturally. 
This is
pictorially depicted in Figure~\ref{fig:gelis1} in terms of the saturation domain.

\begin{figure}
\begin{center}
\includegraphics[width=7.9cm, height=6.1cm]{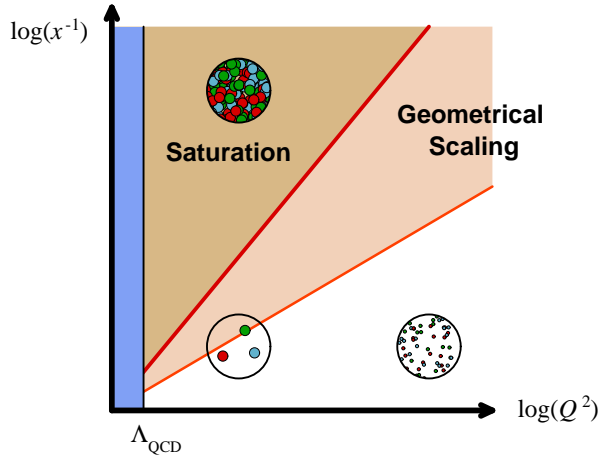}
\end{center}
\caption{\label{fig:gelis1} Saturation domain in the ($Q^2,x$) plane \cite{gelis}.}
\end{figure}

\begin{figure}
\begin{center}
\includegraphics[width=8.1cm, height=7cm]{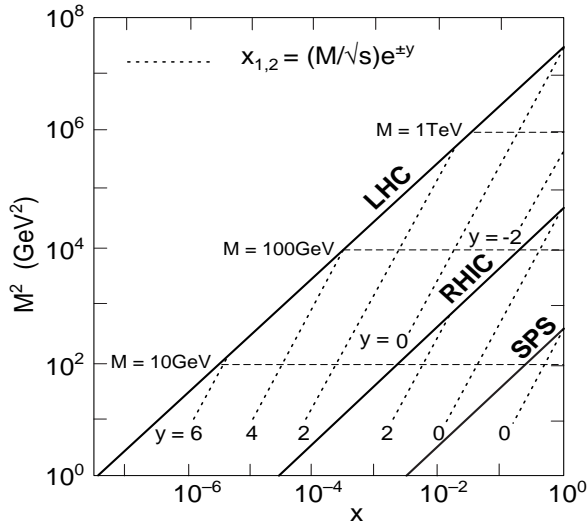}
\caption{\label{fig:xvalues} The range of Bjorken-x values accesible at 
heavy-ion collisions at the top SPS, RHIC and LHC energies \cite{alice-ppr-1}.}
\end{center}
\end{figure}

\begin{figure}
\begin{center}
\includegraphics[width=8.1cm, height=7.9cm]{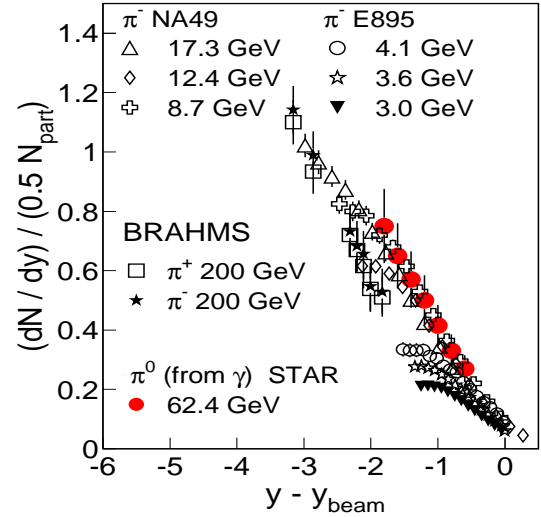}
\caption{\label{fig:star-pmd} Limiting fragmentation as observed from the rapidity 
density distribution for different collision energies \cite{star-pmd1,star-pmd2}. }
\end{center}
\end{figure}

Figure~\ref{fig:xvalues} gives the accessible range of Bjorken-$x$
values and $M^2$ relevant for particle production in nucleus-nucleus collisions at the top
SPS, RHIC and LHC energies~\cite{alice-ppr-1,alice-ppr-2}.
The ALICE experiment will probe a 
continuous range of $x$ as low as about 10$^{-5}$, accessing 
a novel regime where strong nuclear gluon shadowing is expected. 
The study of low $x$ regime, especially at forward rapidities, will be
most appropriate for getting to know the early stage of nuclear collision.

Two recent results from RHIC support the saturation phenomenon: (i) limiting fragmentation, and
(ii) suppression of hadron spectra at froward rapidity.
The limiting fragmentation has been observed~\cite{star-pmd1,star-pmd2} by
combining data for rapidity distributions for several beam energies
and colliding systems. This is shown in Figure~\ref{fig:star-pmd}.
By shifting the rapidity axis by the beam rapidity, one
can see that the distribution for produced particles at collisions of various 
energies and systems tend to an universal curve in the fragmentation region.
This property naturally follows from the CGC framework. From the first measurement of rapidity
distributions at LHC, we will be able to confirm or dispute such a scenario.
The suppression of hadron spectra at forward rapidity in \mbox{d-Au} collisions \cite{bearden}
from the BRAHMS experiment at RHIC is shown in Figure~\ref{fig:gelis3}. The suppression
of the nuclear modification factor at forward rapidities is considered to be a consequence of
the shadowing that builds up via the evolution in rapidity. A new proposal for a forward
electromagnetic calorimeter is being considered which 
will enable ALICE to make detailed study of small $x$ physics.

\begin{figure}
\begin{center}
\includegraphics[width=9cm, height=3.8cm]{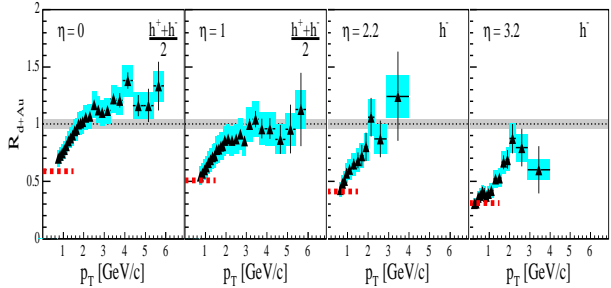}
\caption{\label{fig:gelis3} Forward suppression observed by BRAHMS for d+Au 
collisions at RHIC \cite{bearden}.}
\end{center}
\end{figure}

\section{Summary}

The field of Quark-Gluon Plasma is set to enter a new era with the startup of the LHC. 
The experimental results from  AGS, SPS and RHIC look very promising.
The LHC is set to take this study to a completely new regime where the initial temperature
is supposed to be much larger
and energy density several times higher compared to what has been obtained so far.
Three major experiments, ALICE, ATLAS and CMS will take data during the heavy-ion
collisions of LHC. These three experiments are gearing up to study the new form
of matter in great detail. 
The ALICE detector is designed to cope with the highest particle
multiplicities above those anticipated for \mbox{Pb-Pb} collisions at 
$\sqrt{s_{\rm NN}}=5.5$~TeV. The detector has very large acceptance and excellent
particle identification capabilities. The central barrel measures hadrons, electrons and
photons. The forward muon spectrometer and photon multiplicity detector measure muons and
photons, respectively, at forward rapidities. The Zero Degree Calorimeter, in addition to
providing the beam luminosity for heavy-ion collisions, will also be used for event
characterization. Global observables will provide the bulk properties of the matter in terms
of temperature, energy density, flow and size of the produced fireball. The ALICE experiment
is well suited for precise event-by-event measurements of various observables. The fluctuation
in thermodynamic quantities will provide the nature of the phase transition. 
Formation of DCC domains will signal the chiral symmetry restoration.
The ALICE detector is sensitive to study the domain formation due to DCC.
Transverse momentum distribution of thermal photons and dileptons and their ratios will be measured
in order to gain insight to the space-time evolution of the collision.
The study of
heavy-flavour and quarkonia production will allow for a comparison with QCD predictions and
a study of the in-medium quenching of heavy quarks. The jet rates at the LHC will be quite large.
Measurement of the longitudinal and transverse fragmentation functions of jets will provide
details of the medium properties. Through the study of low-x physics one can also gain insight
to the color glass condensate initial conditions. The ATLAS and CMS detectors complement the
ALICE program by their excellent capabilities for jet and photon measurements at large 
transverse momenta. With the starting of the LHC's heavy-ion program we expect a
host of new fundamental physics within next decade.


\end{document}